\documentclass[aps,prl,twocolumn]{revtex4}
\usepackage{epsfig}
\usepackage{graphicx}
\usepackage{bm, amssymb}
\begin{document}

\title{Controllable-dipole quantum memory}

\author{Khabat Heshami$^{1}$, Adam Green$^{1}$, Yang Han$^{1,2}$, Arnaud Rispe$^1$, Erhan Saglamyurek$^1$, Neil Sinclair$^1$, Wolfgang Tittel$^1$, and Christoph Simon$^1$}
\affiliation{$^1$ Institute for Quantum Information Science and
Department of Physics and Astronomy, University of Calgary,
Calgary T2N 1N4, Alberta, Canada\\
$^2$ College of Science, National University of Defense Technology, Changsha 410073, People's Republic of China}

\begin{abstract}
We present a quantum memory protocol for photons that is based on the direct control of the transition dipole moment. We focus on the case where the light-matter interaction is enhanced by a cavity. We show that the optimal write process (maximizing the storage efficiency) is related to the optimal read process by a reversal of the {\it effective time} $\tau=\int dt g^2(t)/\kappa$, where $g(t)$ is the time-dependent coupling and $\kappa$ is the cavity decay rate. We discuss the implementation of the protocol in a rare-earth ion doped crystal, where an optical transition can be turned on and off by switching a magnetic field.
\end{abstract}
\date{\today}

\maketitle

Quantum memories for light are devices that allow one to store and retrieve light in a way that preserves its quantum state \cite{lvovsky,hammerer,simonEPJD}. They are essential components for optical quantum information processing, notably for quantum repeaters \cite{sangouardRMP}. All quantum memories require a way of switching the coupling between the light and the material system (which is used as the memory) on and off in a controlled way. In the case of memories based on electromagnetically induced transparency or off-resonant Raman transitions \cite{lvovsky,lukinRMP,gorshkovPRL,nunn,Ramanexp} the coupling is controlled by a laser beam, which is typically much more intense than the signal that one aims to store. In contrast, in the case of photon-echo based memories \cite{simonEPJD,tittel,hetet} the coupling is controlled in a more indirect way via the dephasing of the atoms in the storage medium. This typically requires spectral tailoring of the medium by optical pumping before the signal can be stored.

\begin{figure}[b!]
\epsfig{file=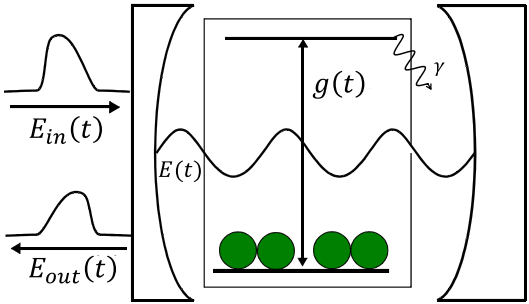,width=0.7 \columnwidth}
\caption{We consider an ensemble of two-level systems inside a one-sided cavity, where the time-dependence of the light-matter coupling $g(t)$ can be controlled. See also Eq. (1).} \label{figure1}
\end{figure}

Here we consider a way of controlling the light-matter interaction that is different from the mentioned examples, and that is particularly simple from a conceptual point of view, namely the direct control of the transition dipole element of the relevant optical transition. This is motivated by recent demonstrations that transition dipoles can be turned on and off in certain solid-state systems, in particular in rare-earth ion doped crystals by applying magnetic fields \cite{Guillot-Noel,LouchetPRB07,GenevaDipole}, and for NV centers in diamond by applying electric fields \cite{Tamarat}.
We consider the case where the storage medium is placed inside an optical cavity \cite{GorshkovCavity,IMQM,vuletic}. This both enhances the light-matter interaction, which is desirable for achieving high efficiencies, and simplifies the equations of motion, thus clearly bringing out the basic principles of the memory dynamics. The free-space case, which is attractive from the point of view of experimental implementation, is discussed in the appendix.

We consider an ensemble of two-level atoms coupled to a cavity mode, see Fig. 1. We ignore the spatial dependence of the light-matter interaction, and thus phase-matching considerations \cite{kalachev,single}. The system that we consider is formally equivalent to a Raman memory in a cavity, if the excited state is adiabatically eliminated in the Raman case \cite{GorshkovCavity}, and where the two-photon spin transition is replaced by a single-photon optical transition.
There is also some similarity to Refs. \cite{Kalachev08, digitalqm}, where the light-matter coupling is controlled by tuning a cavity instead of the transition dipole moment.

We use the usual input-output formalism for a single-sided, fairly high-finesse cavity \cite{scully}. The basic equations are then
\begin{eqnarray}
\dot{\sigma}(t)=-i \Delta(t) \sigma(t) - \gamma \sigma(t) + i g(t) E(t) \nonumber\\
\dot{E}(t)=ig(t) \sigma(t)-\kappa E(t) + \sqrt{2\kappa} E_{in}(t)\nonumber\\
E_{out}(t)=-E_{in}(t)+\sqrt{2\kappa} E(t).
\end{eqnarray}
Thanks to the linearity of the dynamics, $\sigma$ and $E$ can be interpreted as the atomic polarization and cavity fields (in the semi-classical regime), but also as the probability amplitudes corresponding to a single atomic excitation in the ensemble and a single cavity photon respectively (in the quantum regime, which is our focus here) \cite{hammerer,gorshkovPRL,GorshkovCavity}; $E_{in}(t)$ and $E_{out}(t)$ are the incoming and outgoing fields (photon wave functions); $g(t)$ is the time-dependent light-matter coupling, which is proportional to the transition dipole matrix element between the ground and excited atomic states (and also to $\sqrt{N}$, where $N$ is the total number of atoms); $\kappa$ is the cavity decay rate; $\gamma$ is the atomic decay rate; $\Delta(t)$ is a time-dependent detuning, which may arise in practice as a consequence of applying a time-dependent external field in order to control the dipole element and thus $g(t)$; $\gamma$ and $\Delta(t)$ are imperfections that we will neglect at first to keep the discussion simple, but whose effect will be discussed later in the paper.

We are interested in the (realistic) situation where the cavity decay defines the shortest relevant timescale. In this case it is well justified to adiabatically eliminate the cavity field, setting $\dot{E}=0$. This gives
\begin{equation}
E(t)=\frac{1}{\kappa}\left(i g(t) \sigma(t) + \sqrt{2 \kappa} E_{in}(t)\right)
\end{equation}
and hence
\begin{eqnarray}
\dot{\sigma}(t)=-\frac{g^2(t)}{\kappa} \sigma(t)+i \sqrt{\frac{2}{\kappa}} g(t) E_{in}(t) \nonumber\\
E_{out}(t)=E_{in}(t)+i \sqrt{\frac{2}{\kappa}} g(t) \sigma(t)
\label{eom-ae}
\end{eqnarray}
where we have set $\Delta(t)=\gamma=0$, as mentioned above. It is straightforward to derive the (very intuitive) continuity equation
\begin{equation}
\frac{d}{dt} |\sigma(t)|^2=|E_{in}(t)|^2-|E_{out}(t)|^2. \label{cont}
\end{equation}

We now discuss quantum memory operation, starting with a discussion of the read process. (The motivation for this approach will become clear in the following.) The read process corresponds to a situation where there is no incoming photon, $E_{in}=0$. The continuity equation (\ref{cont}) implies
\begin{equation}
|\sigma(0)|^2=|\sigma(t)|^2+\int_0^t dt' |E_{out}(t')|^2,
\end{equation}
which motivates the definition of the read efficiency $\eta_r$ as
\begin{equation}
\eta_r=\frac{\int_0^{\infty} dt |E_{out}(t)|^2}{|\sigma(0)|^2}. \label{etaread}
\end{equation}
Here we have defined $t=0$ as the starting time of the read process.

The solution of Eq. (\ref{eom-ae}) with $E_{in}=0$ is given by
\begin{eqnarray}
\sigma(t)=\sigma(0) e^{-\int_0^t dt' g^2(t')/\kappa} \nonumber\\
E_{out}(t)=i \sqrt{\frac{2}{\kappa}} g(t) \sigma(t).
\label{sigmaread}
\end{eqnarray}
Using Eqs. (\ref{etaread}) and (\ref{sigmaread}) one finds
\begin{equation}
\eta_r=1-e^{-2 \int_0^{\infty} dt g^2(t)/\kappa}. \label{etar}
\end{equation}

Eq. (\ref{etar}) motivates the introduction of the effective time variable
\begin{equation}
\tau=\int_0^t dt' g^2(t')/\kappa, \label{tau}
\end{equation}
 see also Ref. \cite{nunn}, giving the simple expression
$\eta_r=1-e^{-2 \tau_r}$, where $\tau_r=\int_0^{\infty} dt g^2(t)/\kappa$ is the total effective time that elapses during the read process. This means that in order to maximize the read efficiency one simply has to maximize $\tau_r$. The shape of $g(t)$ has an impact on the form of the output field, but the efficiency only depends on $\tau_r$.

In order to rewrite the whole dynamics in terms of the effective time variable $\tau$, we furthermore introduce effective input, output and cavity fields,
\begin{equation}
{\cal E}=\frac{\kappa}{g}E, {\cal E}_{in}=\frac{\sqrt{\kappa}}{g}E_{in},{\cal E}_{out}=\frac{\sqrt{\kappa}}{g} E_{out}. \label{eff-fields}
\end{equation}
One then finds the new equations of motion (after adiabatic elimination of ${\cal E}$)
\begin{eqnarray}
\frac{d}{d\tau} \sigma(\tau)=-\sigma(\tau)+i \sqrt{2} {\cal E}_{in}(\tau) \nonumber\\
{\cal E}_{out}(\tau)={\cal E}_{in}(\tau)+i \sqrt{2} \sigma(\tau).
\label{eom-trf}
\end{eqnarray}
The read efficiency can be rewritten as
\begin{equation}
\eta_r=\frac{\int_0^{\tau_r} d\tau |{\cal E}_{out}(\tau)|^2}{|\sigma(0)|^2}.
\end{equation}
The solution of Eq. (\ref{eom-trf}) in the read case (${\cal E}_{in}=0$) is simply
\begin{equation}
\sigma(\tau)=\sigma(0) e^{-\tau}, {\cal E}_{out}(\tau)=i \sqrt{2} \sigma(\tau).
\label{expo}
\end{equation}
Eq. (\ref{expo}) shows that in terms of the effective time (and of the effective fields) the read process is a simple exponential decay - a remarkable simplification considering that the time dependence of $g(t)$ (and hence $E_{out}(t)$) is completely arbitrary.

We are now ready to discuss the write process. We will immediately use the effective variables. Solving Eq. (\ref{eom-trf}) for non-zero ${\cal E}_{in}$ one finds
\begin{equation}
\sigma(0)=i\sqrt{2} \int_{-\tau_w}^{0} d\tau' e^{\tau'}{\cal E}_{in}(\tau'), \label{sigmatau}
\end{equation}
where $\tau_w$ is the total elapsed effective time for the write process and $\sigma(-\tau_w)=0$. Note that no effective time elapses during times when the transition dipole is zero (i.e. during storage). We define the write efficiency as
\begin{equation}
\eta_w=\frac{|\sigma(0)|^2}{\int_{-\tau_w}^0 d\tau |{\cal E}_{in}(\tau)|^2}.
\end{equation}
Our goal is to find the form of ${\cal E}_{in}(\tau)$ that maximizes $\eta_w$. Since the solution for $\sigma$ is linear in ${\cal E}_{in}$, maximizing $\eta_w$ corresponds to maximizing $|\sigma(\tau_w)|^2$ for a normalized input field satisfying $\int_{-\tau_w}^0 d\tau |{\cal E}_{in}(\tau)|^2=1$.

Before discussing the formal optimization, let us take a step back and try to make a guess for the optimum input field. We have seen that when expressed in terms of effective time rather than real time, the read process simply corresponded to an exponential decay, see Eq. (\ref{expo}). It is natural to suspect that inverting this decay (in effective time) will give the optimum effective input field. This means that our guess for the optimum solution is ${\cal E}_{in}(\tau) \propto e^{\tau}$.

This can be proved by functional differentiation. The optimum solution has to satisfy
\begin{equation}
\frac{\delta}{\delta {\cal E}_{in}^*(\tau)} \left[ |\sigma(0)|^2
+ \lambda \left(\int_{-\tau_w}^0 d\tau |{\cal E}_{in}(\tau)|^2 - 1 \right)  \right]=0,
\end{equation}
where $\lambda$ is a Lagrange multiplier, and ${\cal E}_{in}(\tau)$ and ${\cal E}_{in}^*(\tau)$ are independent variables for each $\tau$. Solving this equation using Eq. (\ref{sigmatau}) gives ${\cal E}_{in}(\tau) \propto e^{\tau}$, confirming the intuitive guess, see also Fig. 2.

For this optimal solution the write efficiency is analogous to the read efficiency,
\begin{equation}
\eta_w=1-e^{-2 \tau_w}.
\end{equation}
The total efficiency (ignoring losses during storage) is then
\begin{equation}
\eta_{tot}=\eta_w \eta_r = (1-e^{-2 \tau_w})(1-e^{-2 \tau_r}),
\end{equation}
which can obviously be simplified further if $\tau_w=\tau_r$.
Provided that the optimum input field is chosen for the write process, the efficiency is thus maximized by maximizing $\tau_w$ and $\tau_r$.

In real time the input field for the write process and the output field for the read process satisfy
\begin{eqnarray}
E_{in}(t) \propto g_w(t) e^{\int_{-\infty}^t dt' g_w^2(t')/\kappa} \nonumber\\
E_{out}(t) \propto g_r(t) e^{-\int_0^t dt' g_r^2(t')/\kappa},
\label{fieldst}
\end{eqnarray}
where $g_w(t)$ and $g_r(t)$ are the light-matter coupling for the write and read processes respectively, and the proportionality constants are such that $\int_{-\infty}^0 dt |E_{in}(t)|^2=1$ and $\int_0^{\infty} dt |E_{out}(t)|^2=\eta_{tot}$. Eq. (\ref{fieldst}) shows that if the light-matter couplings are simple square functions in time, then the input and output fields are growing and declining exponentials in real time, respectively. However, there is no general requirement to choose the couplings in this way. On the one hand, one can achieve optimal write efficiency for any form of $g_w$, as long as the input field satisfies the above equation; on the other hand, the form of the output field can be tailored by choosing the form of $g_r$.

\begin{figure}
\epsfig{file=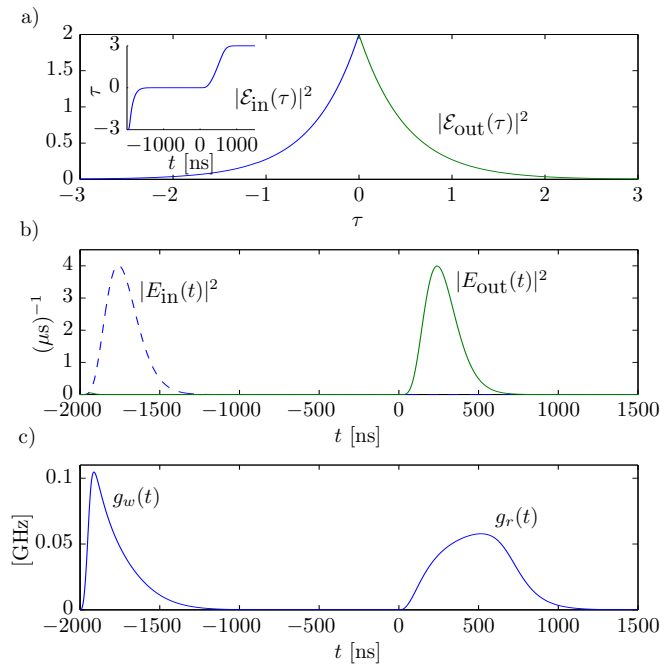,width=\columnwidth}
\caption{(a) The effective input and output fields ${\cal E}_{in}(\tau)$ and ${\cal E}_{out}(\tau)$ of Eq. (\ref{eff-fields}) in terms of the effective time $\tau$ of Eq. (\ref{tau}). The inset shows the effective time versus real time. It can be seen that the effective time elapses only when the coupling is on. (b) An example for the possible time dependence of the real fields $E_{in}(t)$ and $E_{out}(t)$. (c) The corresponding write and read couplings $g_w(t)$ and $g_r(t)$. Any $E_{in}(t)$ can be absorbed with the optimal efficiency $\eta_w=1-e^{-2\tau_w}$ for $g_w(t)$ satisfying Eq. (\ref{couplings}); and $E_{out}(t)$ can, for example, be chosen to be proportional to $E_{in}(t-T)$ (where $T$ is the storage time) for $g_r(t)$ satisfying Eq. (\ref{couplings}).} \label{figure2}
\end{figure}

This means in particular that memory performance can be optimal even if the input and output fields are not related by time reversal in real time. For example, let us suppose that we want the input and output fields to have the same temporal shape, $E_{out}(t)=-\sqrt{\eta_w \eta_r} E_{in}(t-T)$, where $T$ is the storage time, while still satisfying Eq. (\ref{fieldst}). By inverting Eq. (\ref{fieldst}) one can show that this can be achieved by choosing the following time-dependent couplings for the write and read processes:
\begin{eqnarray}
g_w(t)=\sqrt{\frac{\kappa \eta_w |E_{in}(t)|^2}{2(1-\eta_w+\eta_w \int_{-\infty}^t dt'|E_{in}(t')|^2)}} \nonumber\\
g_r(t)=\sqrt{\frac{\kappa \eta_r |E_{in}(t-T)|^2}{2(1-\eta_r \int_{-\infty}^t dt'|E_{in}(t'-T)|^2)}} \label{couplings}.
\end{eqnarray}
This choice of $g_w(t)$ achieves the optimal write efficiency $\eta_w=1-e^{-2\tau_w}$ for any input field $E_{in}(t)$ and any value of $\tau_w=\int_{-\infty}^0 dt g_w^2(t)/\kappa$. On the other hand, the above choice of $g_r(t)$ ensures that the output field is proportional to the input field (shifted in time by $T$). We have seen that the read efficiency always satisfies
 $\eta_r=1-e^{-2\tau_r}$ with $\tau_r=\int_0^{\infty} dt g_r^2(t)/\kappa$. Note that arbitrary output field shapes are possible for appropriately chosen $g_r(t)$, see also Fig. 2.

So far we have neglected the spontaneous decay rate $\gamma$. It is not difficult to include in the above approach, but it obviously leads to somewhat lower efficiencies, because its effect is irreversible. The optimum input field can still be found by functional differentiation. To discuss the simplest example, let us consider square coupling pulses of strength $g_{w(r)}$ and duration $t_{w(r)}$. Then the optimized input field for writing satisfies
$E_{in}(t) \propto g_w e^{\frac{g_w^2 t}{\kappa}+\gamma t}$
and the output field from the read process fulfills
$E_{out}(t) \propto g_r e^{-\frac{g_r^2 t}{\kappa}-\gamma t}$,
while the efficiencies satisfy
\begin{equation}
\eta_{w(r)}=\frac{\frac{g_{w(r)}^2}{\kappa}}{\frac{g_{w(r)}^2}{\kappa}+\gamma}
\left(1-e^{-2\left(\frac{g_{w(r)}^2}{\kappa}+\gamma\right) t_{w(r)}} \right). \label{eff-gamma} \end{equation}
One can see that for large effective times the efficiencies tend towards $\frac{C}{C+1}$ \cite{GorshkovCavity}, where $C=\frac{g^2}{\kappa \gamma}$, which is essentially the optical depth in the presence of the cavity. High efficiencies require large $C$. For a given decay rate, $C$ can in principle always be increased by increasing $g$ (which requires increasing the dipole moment or the number of atoms), or by decreasing $\kappa$ (which requires increasing the finesse of the cavity, i.e. the number of roundtrips).

The general case also includes a time-dependent detuning $\Delta(t)$. By functional differentiation one finds that the optimum input field has a phase dependence that exactly compensates the detuning. If this is not possible, the achievable efficiencies will again be reduced. However, in analogy to the case of spontaneous decay, the effect will be small as long as the ratio $\frac{g^2}{\kappa \Delta}$ is large.

We will now discuss potential experimental implementations of the proposed protocol. In certain rare-earth ion doped crystals optical transitions can be switched on and off by changing the applied magnetic field \cite{Guillot-Noel,LouchetPRB07,GenevaDipole}. This is due to the coupling of the electronic Zeeman and hyperfine interactions in the presence of the crystal field. This coupling yields a substantial contribution to the overall nuclear Zeeman effect which is different for the ground and excited states, allowing one to control the branching ratios of optical transitions. For example, in Tm:YAG adding a field of order 80 mT transversally to a static applied field of 1 T will turn on a previously forbidden transition to a point where its optical depth $d$ is of order 1/cm \cite{Guillot-Noel,LouchetPRB07}. It is possible to control magnetic fields of this order (tens of mT) on ns timescales \cite{Bswitching}, making it possible to store light pulses whose duration is on these timescales. See also the appendix for more details on the proposed implementation. In practice the spectral width of the pulses is more likely to be limited by nearby transitions.
The optical depth will be enhanced by the cavity, one has $C \approx d F$ for the ratio $C$ defined above, where $F$ is the cavity finesse. Based on Eq. (\ref{eff-gamma}) high efficiencies should thus be achievable combining crystals of typical dimensions (say 1 cm in length) with moderate-finesse cavities. We have focused on the case of a memory inside a cavity. However, good memory performance based on the same principle is possible without a cavity as well, see the appendix. In particular, we show that the present protocol outperforms memories based on controlled reversible inhomogeneous broadening (CRIB) \cite{kraus} in terms of efficiency for a given optical depth.

The described memory could be attractive from a practical point of view as a solid-state Raman-like memory that does not require an optical control field, thus avoiding spurious signal detections (i.e. noise) due to the presence of the strong control beam \cite{Ramanexp}. Implementations in systems other than rare-earth ion doped crystals may be possible, for example using electric control fields for NV centers in diamond \cite{Tamarat}.

More conceptually, the present protocol has the potential to provide insight into the basic principles underlying quantum memories for light in general. As a first example, we have seen that the optimal write process is related to the read process by a reversal of effective, but not necessarily real, time. Because of the mentioned formal equivalence of the considered system to off-resonant Raman memories, this result applies to the latter as well. It is an interesting question whether the same also holds for other memory protocols for appropriately defined effective variables. See Refs. \cite{gorshkovPRL,kraus,gorshkovII,moiseev} for related discussions in real time. Even more generally, the present protocol seems well placed to serve as an ``archetype'' for quantum memories, because, as discussed above, in all memory protocols the light-matter interaction is controlled in some fashion. Mapping various protocols onto the controllable-dipole memory discussed here may be a good way of analyzing their similarities and differences.

{\it Acknowledgment} - We thank Josh Nunn and Daniel Oblak for useful discussions. This work was supported by AITF, NSERC, the China Scholarship Council, General Dynamics Canada, and iCORE (now part of Alberta Innovates).

\appendix
\section{Appendix A1: Solution of the Maxwell-Bloch equations in free space}

The equations of motion in free space are given by
\begin{eqnarray}\label{Max-Bloch}
\nonumber \partial_t \sigma(z,t) &=& -(\gamma+i\Delta(t))\sigma(z,t) + i g(t) E(z,t)  \\
\partial_z E(z,t) &=& i \frac{g(t)}{c} \sigma(z,t),
\end{eqnarray}
where $\sigma(z,t)$ can be interpreted as the wave function for a single atomic excitation, and $E(z,t)$ as the wave function for a single photon; $\gamma$, $\Delta(t)$ and $g(t)$ are the same as in the paper; the latter is given by $g(t)\equiv \sqrt{\omega_0/2\epsilon_0\hbar V}\wp(t)$, where $\wp(t)$ is the controllable transition dipole moment, $\omega_0$ is central frequency of the incident light pulse, and $V$ is the quantization volume; $c$ is the speed of light. These equations are valid if saturation can be neglected, which is guaranteed if the number of atoms $N$ is much greater than one. Moreover it is assumed in the derivation that the length of the medium $L$ is much smaller than the characteristic length of the pulse, such that the difference between the real time $t$ and retarded time $t-z/c$ is negligible.

We define $S(z,t)=e^{i\chi(t)}\sigma(z,t)$, ${\cal E}(z,t)=\frac{c E(z,t)}{ig(t)}e^{i\chi(t)}$ and $\tau(t)=\int_{0}^{t}dt'\frac{g(t')^2}{c}$, where $\chi(t)=\int_{0}^{t}dt'(\Delta(t')-i\gamma)$, see also Ref. \cite{nunn}. Substituting these variables in the Eq. (\ref{Max-Bloch}) gives
\begin{eqnarray}\label{Max-Bloch-simp}
\nonumber \partial_{\tau} S(z,\tau) &=& -{\cal E}(z,\tau)  \\
\partial_z {\cal E}(z,\tau) &=& S(z,\tau).
\end{eqnarray}
Using a proper Laplace transformation, ${\cal L}\{ {\cal E}(z,\tau)\} =e(s,\tau)=\int_0^{\infty}dz e^{-sz}{\cal E}(z,\tau)$ and ${\tilde S}(s,\tau)=\int_0^{\infty}dz e^{-sz} S(z,\tau)$, one can convert the set of differential equations in Eq. (\ref{Max-Bloch-simp}) to a differential equation and an algebraic equation, namely
\begin{eqnarray}\label{Max-Bloch-simp-lap}
\nonumber \partial_{\tau} {\tilde S}(s,\tau) &=& -e(s,\tau)  \\
{\tilde S}(s,\tau) &=& -{\cal E}(0,\tau)+s e(s,\tau).
\end{eqnarray}
One can easily find ${\tilde S}(s,\tau)=-\frac{1}{s}\int_{0}^{\tau} d{\tau '} e^{(\tau ' -\tau)/s} {\cal E}(0,\tau ')+e^{-\tau/s}{\tilde S}(s,0)$. Plugging this result into the second equation in Eq. (\ref{Max-Bloch-simp-lap}) gives $e(s,\tau)=s^{-1}{\cal E}(0,\tau)+s^{-1} e^{-\tau/s} {\tilde S}(s,0) - s^{-2} \int_{0}^{\tau} d{\tau '} e^{(\tau ' -\tau)/s} {\cal E}(0,\tau ')$. Using ${\cal L}\{ (z/a)^{n/2} I_n(\sqrt{4az})\} =s^{-(n+1)}e^{a/s}$ and the convolution theorem \cite{convolution} the solution is as follows,
\begin{eqnarray}\label{epsilon-solution}
{\cal E}(z,\tau)&=&{\cal E}(0,\tau) + \int_{0}^{z}dz' S(z',0) I_0(\sqrt{4\tau(z'-z)}) \\ \nonumber
&+& \int_{0}^{\tau} d{\tau}' {\cal E}(0,\tau ') \sqrt{\frac{z}{{\tau}'-\tau}} I_1(\sqrt{4({\tau}'-\tau)z}),\\ \nonumber
S(z,\tau)&=&S(z,0)-\int_{0}^{\tau}d{\tau '} {\cal E}(0,\tau ') I_0(\sqrt{4z(\tau ' -\tau)})\\ \nonumber
&-& \int_{0}^{z} d{z'} S(z',0) \sqrt{\frac{\tau}{z'-z}} I_1(\sqrt{4\tau(z'-z)}),
\end{eqnarray}
where $I_n$ is the $n$th modified Bessel function, see also \cite{Raymer}. One can simply use the above-mentioned definitions for ${\cal E}(z,\tau)$, $S(z,\tau)$ and $\tau(t)$ to convert the result in Eq. (\ref{epsilon-solution}) to the actual optical field.

This solution allows one to analyze both the storage (write) and retrieval (read) process, depending on the initial conditions. In the next section it is used to determine the efficiency of the controllable-dipole memory in free space.

%

\section{Appendix A2: Efficiency analysis and comparison with the controlled reversible inhomogeneous broadening quantum memory protocol}

\begin{figure}
\epsfig{file=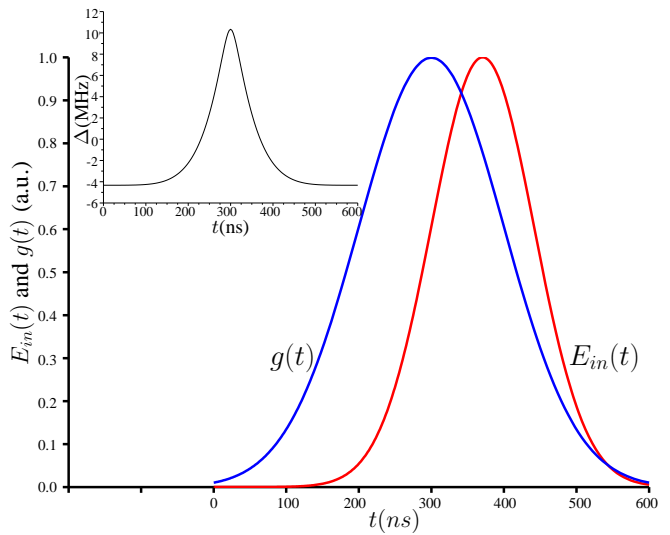,width=\columnwidth}
\caption{\label{fig1} The time-dependent coupling $g(t)$ and input field $E_{in}(t)$ used in our examples. The inset shows the time-dependent detuning $\Delta(t)$. See text for a more detailed discussion.}
\label{gt-Ein-Delta}
\end{figure}

In this section we study the efficiency of the controllable-dipole quantum memory and we compare its performance to that of the controlled reversible inhomogeneous broadening (CRIB) protocol \cite{CRIB}. In the free-space case, in contrast to the cavity case discussed in the paper, we do not know the exact conditions for $E_{in}(t)$ and $g(t)$ which maximize the total efficiency for the controllable-dipole protocol. In spite of this, we show that it is still possible to achieve very good memory performance under realistic conditions. We have chosen an incident light pulse that is a Gaussian with full width at tenth of maximum (FWTM) $T_{pulse}=300$ ns, and a Gaussian profile for $g(t)$, which is displaced in time relative to the pulse, see Fig. \ref{gt-Ein-Delta}. We also choose $\gamma=50$ kHz, and assume that $\Delta(t)$ has the form shown in the inset of Fig. \ref{gt-Ein-Delta}. All of these choices are motivated by the implementation considerations discussed in the next section. We are interested in the efficiency of the memory as a function of the optical depth.  In our model, the optical
depth $D(t)=\frac{g(t)^2 L}{\gamma c}$ is a function of time. In order to facilitate the comparison with CRIB we define $d=\max D(t)$ to indicate the effective optical depth of our system when the dipole has its maximum value.

\begin{figure}[t]
\epsfig{file=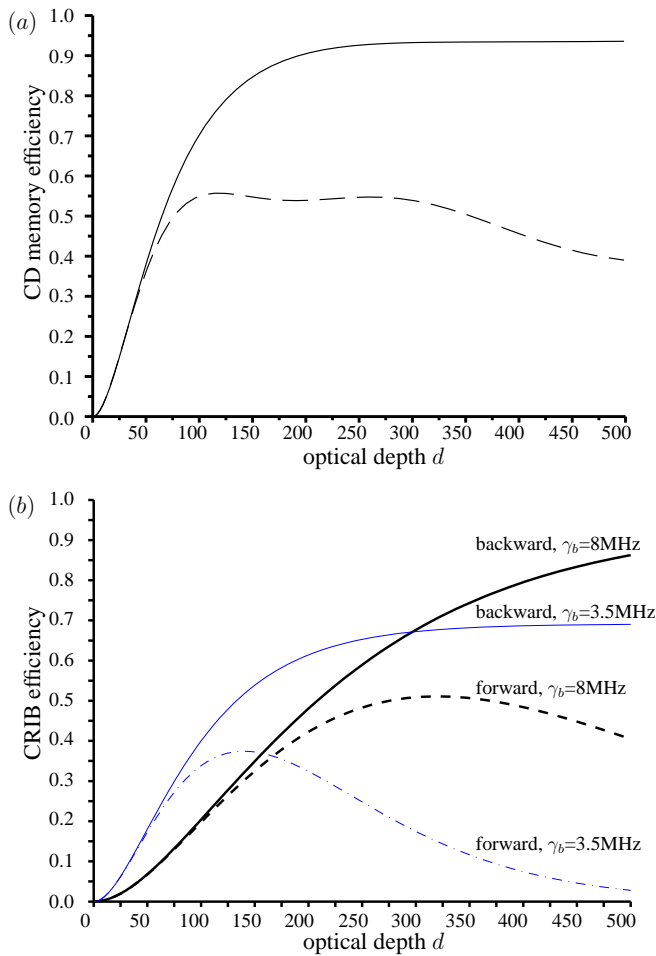,width=\columnwidth}
\caption{\label{fig3} (a) Efficiency of the controllable-dipole memory in free space for readout in the backward (solid line) and forward (dotted line) direction. In analogy with other quantum memory protocols such as AFC or CRIB, the forward retrieval efficiency is limited due to re-absorption. (b) Efficiency for a CRIB memories with two different values for the broadened linewidth, also in backward and forward direction. See text for a more detailed discussion.
\label{efficiency}}
\end{figure}

Fig. \ref{efficiency}(a) shows the efficiency of the controllable-dipole memory for the above-mentioned parameters as a function of $d$, for retrieval in the backward and in the forward direction. As for other memory protocols including CRIB, the efficiency in the forward direction is limited by re-absorption, see Ref. \cite{Sangouard}. In the backward direction, the achievable maximum efficiency is mainly limited by the decay rate $\gamma$. Note that backward retrieval requires transferring the atomic excitation to an extra level using an optical control field. Here we have assumed that the pulse is retrieved immediately after having been stored. Otherwise the storage time also has to be taken into account.

Fig. \ref{efficiency}(b) shows the efficiency as a function of $d$ for the CRIB memory protocol, based on the results of Ref. \cite{Sangouard}. In order to make a meaningful comparison, we choose the same initial linewidth $\gamma=50$ kHz for CRIB as for the controllable-dipole memory. In the CRIB protocol the initial line is broadened through the application of an external field in order to accomodate the spectrum of the input pulse. This can be seen as analogous to the effective spectral broadening that happens in the controllable-dipole protocol as a consequence of the time dependence of the transition dipole, making CRIB a natural point of comparison for the present protocol. The input pulse of Fig. \ref{gt-Ein-Delta} has a frequency FWTM of 9.8 MHz. Since broadening the initial line lowers the optical depth, it is advantageous to choose the width of the broadened line somewhat smaller than this value. This cuts off the outermost frequency components of the pulse, but enhances the efficiency for the most important components. In Fig. \ref{efficiency}(b) we show the efficiency of CRIB for two different choices of the broadened width. The first (8 MHz)  is chosen such that the maximum achievable efficiency is the same as for the controllable-dipole memory. One can see that in this case the efficiency for the CRIB memory increases significantly more slowly with $d$. On the other hand, for the second choice of broadened linewidth (3.5 MHz) the efficiency initially increases similarly quickly for CRIB. But then the achievable maximum efficiency is reduced, because a significant fraction of the input pulse is cut off. Taking these observations together, one can see that the controllable-dipole memory shows better efficiency performance than the CRIB protocol, even without full optimization of the pulse shapes. It should be noted that the shape of the broadened absorption line in CRIB has not been optimized for a Gaussian pulse.

\section{Appendix A3: Implementation in Tm:YAG}

\begin{figure}
\epsfig{file=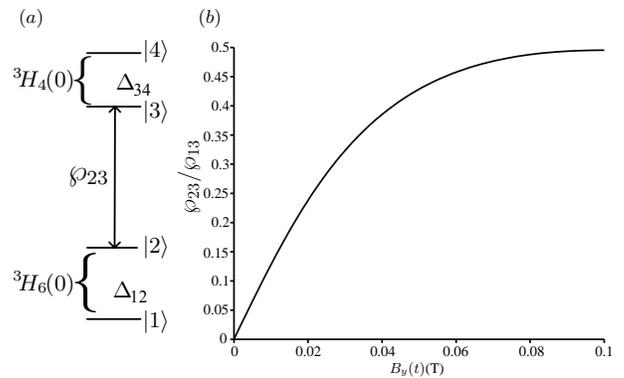,width=\columnwidth}
\caption{(Color online)(a) Transition used for the proposed implementation. (b) The corresponding transition dipole moment as a function of the applied magnetic field. See text for a more detailed discussion.}
\label{transition-dipole}
\end{figure}

\begin{figure}
\epsfig{file=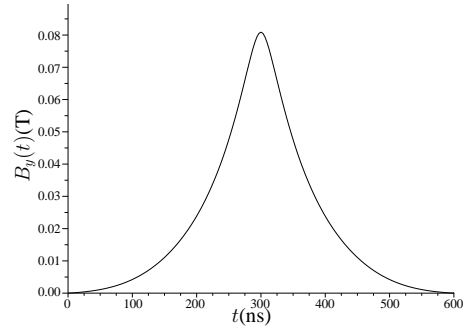,width=0.7 \columnwidth}
\caption{\label{B-field} Time dependence of the applied magnetic field $B_y$ in order to obtain $g(t)$ as shown in Fig. \ref{gt-Ein-Delta}; this field dependence also leads to a time-dependent detuning $\Delta(t)$ as shown in the inset of Fig. \ref{gt-Ein-Delta}.}
\end{figure}

In the following we discuss a potential implementation of the controllable-dipole quantum memory protocol in a Tm:YAG crystal. We focus on the first crystal-field states of the $^{3}H_{6}(0)$ and $^{3}H_{4}(0)$ multiplets, with a transition at 793 nm \cite{LouchetPRB07}, see Fig. \ref{transition-dipole}(a). Using the same definition of local crystal-field axes as in Ref.\cite{LouchetPRB07}, for a magnetic field in $x$ direction the states $|1\rangle, |2\rangle, |3\rangle,|4\rangle$ are eigenstates of the same nuclear spin projection, where $|1\rangle$ and $|3\rangle$ correspond to $M_I=1/2$ and $|2\rangle$ and $|4\rangle$ correspond to $M_I=-1/2$ respectively. In this case the transition from $|2\rangle$ to $|3\rangle$ is forbidden by the nuclear spin selection rule, corresponding to $g(t)=0$. However, when the direction of the magnetic field is changed, the nuclear spin eigenstates evolve differently for the ground ($|1\rangle,|2\rangle$) and excited ($|3\rangle,|4\rangle$) states. This is due to the cross coupling of the electronic Zeeman effect and the hyperfine interaction, see \cite{Guillot-Noel}. As a consequence, there can be a considerable transition dipole moment between $|2\rangle$ and $|3\rangle$. Based on the crystal-field Hamiltonian approach, one can completely calculate the magnetic interactions from the crystal-field wave functions \cite{Guillot-Noel}. Fig.\ref{transition-dipole}(b) shows a particular case of the dependence of $\wp_{23}$ on the direction of $B$. We fix $B_{x}=1T, B_{z}=0T$, and let $B_{y}$ vary from 0T to 0.10T. As a consequence, $\wp_{23}$ varies from 0 to 0.44$\wp_{13_0}$, where $\wp_{13_0}\equiv \wp_{13}|_{(B_y=0)}$ is the transition dipole moment of $|1\rangle\leftrightharpoons|3\rangle$ when $|B_y|=0$. Note that $\wp_{13}$ also varies slightly under the same modulation of the direction of the magnetic field, see Fig. \ref{transition-dipole}. Therefore, with a magneto-dependent transition dipole moment and long-lived upper level (coherence times of order 100 $\mu$s have been reported \cite{coherence}), the transition $|2\rangle \leftrightharpoons |3\rangle$ is an excellent candidate for the present scheme. The initial narrow line can be prepared by spectral tailoring, see e.g. \cite{tailoring}.

The speed of controlling $g(t)$ is limited by how fast one can control the magnetic field. Ref. \cite{Bswitching} demonstrated a device  composed of an electronic circuit and a low inductance coil capable of producing rapidly switched magnetic fields with a speed of 0.02T/10ns. Noticing the variation range of $B_{y}$ in Fig.\ref{transition-dipole}(b), this speed would enable us to control $\wp_{23}$, and consequently change the coupling $g(t)$, on a time scale of 100ns, which sets a lower bound for the input pulse duration $T_{pulse}$. This motivates the choice of $g(t)$ and $E_{in}(t)$ shown in Fig. \ref{gt-Ein-Delta}. Fig.\ref{B-field} shows the time-dependent magnetic field $B_{y}$ that has to be applied in order to produce that form of $g(t)$. This field also leads to a time-dependent detuning $\Delta(t)$ as shown in the inset of Fig. \ref{gt-Ein-Delta}.

\end{document}